\documentclass{aa}

\usepackage{graphics}
\usepackage{psfig}

\begin{document}

\thesaurus{04
 (11.09.1 NGC 4418;  
  11.09.4;           
  11.19.1;           
  13.09.4)}          

\title{The obscured mid-infrared continuum of NGC\,4418:\\
       a dust- and ice-enshrouded AGN
       \thanks{Based on observations with ISO, an ESA project with 
       instruments funded by ESA Member States (especially the PI 
       countries: France, Germany, the Netherlands and the United 
       Kingdom) and with the participation of ISAS and NASA}}

\author{H.W.W. Spoon\inst{1}     \and 
        J.V. Keane\inst{2}       \and 
        A.G.G.M. Tielens\inst{2} \and 
        D. Lutz\inst{3}          \and 
        A.F.M. Moorwood\inst{1}}

\offprints{H.W.W. Spoon}

\institute{European Southern Observatory,
           Karl-Schwarzschild Strasse 2, D-85748 Garching, Germany\\
           email: hspoon@eso.org,amoor@eso.org
       \and
           Kapteyn Astronomical Institute, Postbus 800, 
           NL-9700 AV Groningen, The Netherlands\\
           email: j.keane@astro.rug.nl, tielens@astro.rug.nl
       \and
           Max-Planck-Institut f\"ur Extraterrestrische Physik (MPE),
           Postfach 1312, D-85741 Garching, Germany\\
           email: lutz@mpe.mpg.de}

\date{}

\maketitle

\begin{abstract}
We report the detection of absorption features in the 6--8\,$\mu$m 
region superimposed on a featureless mid-infrared continuum in
NGC\,4418. For several of these features this is the first detection 
in an external galaxy. We compare the absorption spectrum of NGC\,4418
to that of embedded massive protostars and the Galactic centre,
and attribute the absorption features to ice grains and to 
hydrogenated amorphous carbon grains. From the depth of the ice 
features, the powerful central source responsible for the mid-infrared
emission must be deeply enshrouded. 
Since this emission is warm and originates in a compact region,
an AGN must be hiding in the nucleus of NGC\,4418.

\end{abstract}

\section{Introduction}
\begin{figure}
\resizebox{\hsize}{!}{\includegraphics{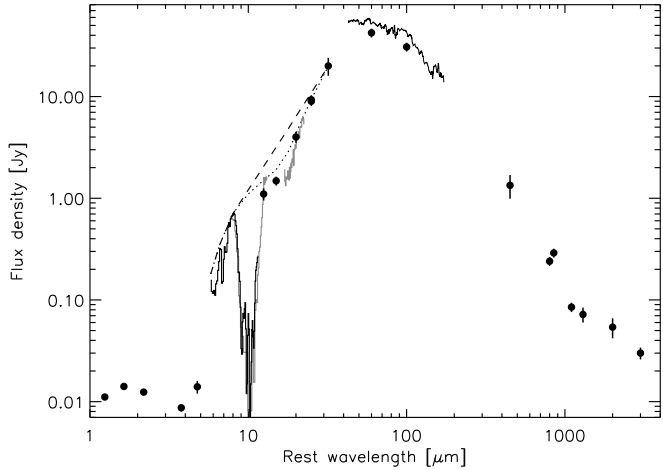}}
\caption{The 1-3000\,$\mu$m spectral energy distribution of NGC\,4418,
including the new ISO-PHT-S, ISO-CAM and smoothed ISO-LWS data. The 
data shown in grey is the spectrophotometry of Roche et al. (\cite{Roche86}). 
Other data shown has been taken from Dudley (\cite{Dudley97}),
Lisenfeld et al. (\cite{Lisenfeld}), Roche \& Chandler (\cite{Roche93}),
Soifer et al. (\cite{Soifer}) and Wynn-Williams \& Becklin (\cite{Wynn}).
The dotted and dashed curves show two choices for the local mid-IR 
continuum. The dotted line interpolates the peaks of the observed
data, whereas the dashed line assumes a stronger 18\,$\mu$m silicate 
absorption.}
\label{mirfir_sed}
\end{figure}

The ISO mission has considerably
enhanced our knowledge of the mid-IR properties of normal, starburst, 
Seyfert and Ultra-luminous Infrared Galaxies (ULIRGs). The spectra 
of most sources are dominated by ISM emission features, 
the most prominent of which are the well-known PAH emission bands
at 6.2, 7.7, 8.6, 11.3 and 12.7\,$\mu$m. 
The PAH features and the emission lines have been used qualitatively
and quantitatively as diagnostics for the ultimate physical processes
powering the galactic nuclei (Genzel et al. \cite{Genzel}; Lutz
et al. \cite{Lutz}; Rigopoulou et al. \cite{Rigopoulou}; Tran et al.
\cite{Tran}). A broad absorption band due to the Si-O stretching mode 
in amorphous silicates, centered at 9.7\,$\mu$m, is also commonly 
detected in galaxies.
Since the center of the silicate absorption coincides with a gap 
between the 6.2--8.6\,$\mu$m and 11.3--12.8\,$\mu$m PAH complexes, 
it is not readily apparent whether a 9.7\,$\mu$m flux minimum should
be interpreted as the ``trough'' between PAH emission features or as 
strong silicate absorption, or as a combination of the two.
In spectra observed towards heavily extincted Galactic lines of 
sight, a strong silicate feature is often accompanied by ice
absorption features in the 6--8\,$\mu$m region (e.g. Whittet et al.
\cite{Whittet}).
Until recently this combination had not been reported in 
equally extincted extragalactic sources, despite detections of
ice features at shorter wavelengths (Spoon et al. \cite{Spoon};
Sturm et al. \cite{Sturm}). In this {\it Letter} we report on 
the detection of ices in the strongly absorbed, ISO-PHT-S spectrum 
of NGC\,4418, a nearby (D=29 Mpc; 1$\arcsec$=140pc) luminous 
(L$_{\rm IR}$=10$^{11}$\,L$_{\odot}$) bright IRAS galaxy. NGC\,4418
and the distant ULIRG IRAS\,00183-7111 (Tran et al. \cite{Tran}) are 
the first detections of these ice features in external galaxies.\\
\indent NGC\,4418 is well-known for its deep 9.7\,$\mu$m silicate 
feature (Roche et al. \cite{Roche86}). Additional evidence for
strong extinction is the weakness (H$\alpha$) and absence 
(H$\beta$, Br$\alpha$, Br$\gamma$) of hydrogen recombination line
emission (Kawara et al. \cite{Kawara89}; Ridgway et al. 
\cite{Ridgway}; Lehnert \& Heckman \cite{Lehnert}; L.Kewley, 
priv.comm.), commonly detected in galaxies.
HST-NICMOS images (Scoville et al. \cite{Scoville}) show hardly 
any structure in the inner 400pc$\times$400pc, except for large 
scale extinction ($\Delta$A$_{\rm V}\sim$2). 
The IRAS colors of NGC\,4418 indicate that --- unlike most other 
galaxies --- the 12--100$\mu$m emission of NGC\,4418 is dominated 
by a warm dust component, peaking shortward of 60$\mu$m (see 
Fig.\,\ref{mirfir_sed}).
VLA radio maps at 6 and 20\,cm (Condon et al. \cite{Condon}; Eales 
et al. \cite{Eales}) show NGC\,4418 to be compact (70pc$\times$50pc
at most). This, as well as the presence of large quantities 
of warm dust, has been taken as evidence for the presence of an 
otherwise hidden AGN in NGC\,4418.
In this {\it Letter} we present mid-IR spectral evidence 
lending further support for the presence of an AGN in NGC\,4418.

\vspace*{-5mm}

\section{Observations}

A low resolution ($\lambda$/$\Delta\lambda\sim$90) ISO-PHT-S spectrum
of the central 24$\arcsec\times$24$\arcsec$ of NGC\,4418 was obtained 
on 1996 July 14 as part of a project on the interstellar medium of 
normal galaxies (Helou et al. \cite{Helou}).
The measurement was carried out in triangular chopped mode, using
a chopper throw of 150$\arcsec$. The resulting spectrum is thus free of
contributions from zodiacal light. The ISO-PHT-S data were reduced using 
PIA 8.2.
The absolute calibration is accurate to within 20\%. 
The resulting spectrum is shown in Fig.\,\ref{mir_seds}.\\
\indent In Fig.\,\ref{mirfir_sed} we compile the 1--3000\,$\mu$m 
(spectro)photometric observations of NGC\,4418, including the new,
standard reduced, ISO-PHT-S, ISO-CAM and 
ISO-LWS observations. Stellar emission probably dominates the near-IR 
regime up to 4$\mu$m, beyond which a strong mid-infrared continuum
sets in. The SED reveals that a substantial part of the emission
emanates from optically thick warm dust (Roche \& Chandler \cite{Roche93}; 
Lisenfeld et al. \cite{Lisenfeld}), peaking at 40--60\,$\mu$m. Also shown 
in Fig.\,\ref{mirfir_sed} are two equally possible choices of local 
continuum which we have adopted to analyse the ice, hydrogenated 
amorphous carbon (HAC) and silicate absorption. This will be discussed
in detail in Section 3.2.

\section{The mid-IR spectrum of NGC\,4418}

\subsection{Spectral features}

\begin{figure}
\resizebox{\hsize}{!}{\includegraphics{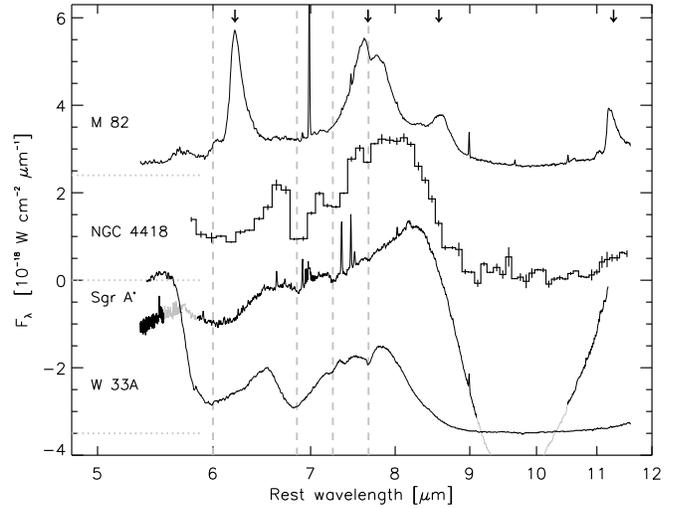}}
\caption{A comparison of the ISO spectra of NGC\,4418, M\,82, 
the embedded massive protostar W\,33A and the Galactic centre 
(Sgr\,A$^{*}$). Except for NGC\,4418, the spectra have been scaled and
offset. We removed the strong 7$\mu$m $[$\ion{Ar}{ii}$]$ line from the
Sgr\,A$^*$ spectrum. The vertical dashed lines facilitate comparision 
between the four spectra with well known Galactic absorption features. 
The arrows mark the rest wavelengths of the 6.2, 7.7, 8.6 and 11.3$\mu$m 
PAH features. The zero flux levels for M\,82, NGC\,4418 and W\,33A
are indicated with horizontal dotted lines.}
\label{mir_seds}
\end{figure}

The mid-IR spectrum of NGC\,4418 (Fig.\,\ref{mir_seds}) bears little 
resemblance to the spectrum of (almost) any other galaxy obtained by ISO. 
The spectra of normal and starburst galaxies (Rigopoulou et al. 
\cite{Rigopoulou}; Helou et al. \cite{Helou}) are dominated by strong 
PAH emission features. Seyfert galaxies with a clear line of sight to 
AGN-heated hot dust, on the other hand, are dominated by a strong mid-IR 
continuum with PAHs barely recognizable (Clavel et al. \cite{Clavel}). 
The mid-IR spectrum of NGC\,4418 is rich in features but does not
resemble PAH spectra. Rather it is similar to spectra observed towards 
heavily extincted Galactic lines of sight, such as deeply embedded massive 
protostars or the Galactic centre (Fig.\,\ref{mir_seds}). These show 
no evidence for PAH emission features but do show strong absorption 
features. A simple criterion for the role of PAH emission and absorption
features is based on the location of maxima in the 6--7\,$\mu$m region:
PAH spectra show the 6.2\,$\mu$m emission feature, whereas absorption
spectra show a peak at 6.5--6.7\,$\mu$m, which is not an emission
feature but a window of reduced absorption between two features.

\begin{figure}
\resizebox{\hsize}{!}{\includegraphics{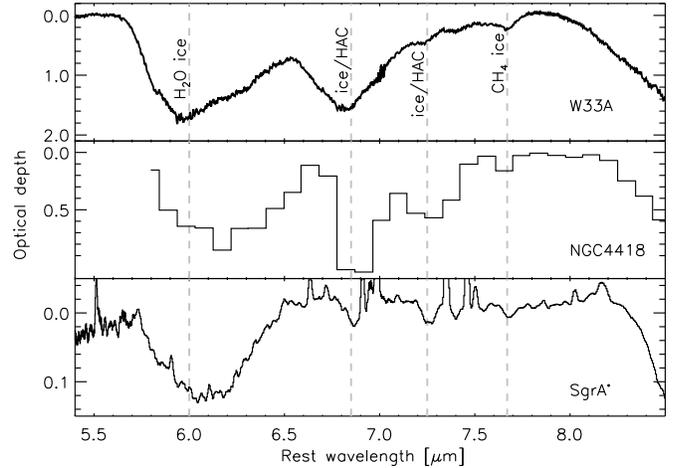}}
\caption{The optical depth spectra for W\,33A, NGC\,4418, and 
Sgr\,A$^{*}$. The vertical lines indicate the positions of the 6.0, 
6.85, 7.3, and 7.6\,$\mu$m ice and HAC absorption bands seen toward 
Galactic lines of sight.}
\label{ice_depths}
\end{figure}

A detailed comparison of the absorption features with Galactic templates 
may be able to shed further light on the origin of the extinction in 
NGC\,4418. The absorption features were determined by fitting a local 
polynomial to the peaks of the 5.8--8\,$\mu$m spectrum of NGC\,4418. 
The absence of the 6.2\,$\mu$m PAH feature indicates that emission
features are very weak in the spectrum, and justifies this simple 
procedure to derive the shape and depth of the absorption features.
Fig.\,\ref{ice_depths} shows the resulting NGC\,4418 optical depth 
spectrum along with the optical depth spectrum of ices towards the
massive protostar W\,33A (Gibb et al. \cite{Gibb}) and the spectrum of 
the Galactic centre (Sgr\,A$^{*}$), which displays absorptions due 
to ices as well as features due to dust in the diffuse ISM 
(Chiar et al. \cite{Chiar}). The spectrum of NGC\,4418 shows 
absorption features at 6.0, 6.8, 7.3, 7.6, 10 and 18\,$\mu$m 
(Table\,\ref{optdepths}). The 6.0\,$\mu$m feature in NGC\,4418 is
similar to that in the Galactic sources but with a perhaps more 
pronounced long wavelength wing. The 6.85\,$\mu$m feature is
considerably narrower than the molecular cloud feature but is similar
in width to the diffuse ISM feature. The band at 7.3\,$\mu$m
is substantially broader than the molecular cloud and diffuse
ISM features. The absorption band near 7.6\,$\mu$m is similar
to that observed locally.\\
\indent The presence of ice along the line of sight toward NGC\,4418 is
suggested by the identification of the 7.6\,$\mu$m band with CH$_{4}$ 
(Boogert et al. \cite{Boogert96}, \cite{Boogert97}) and by the presence
of the 6.0\,$\mu$m band due to H$_{2}$O ice (Keane et al. \cite{Keane}; 
Chiar et al. \cite{Chiar}). The origins of the 6.85\,$\mu$m and 7.3\,$\mu$m
bands in NGC\,4418 are unclear. Interstellar ices also show 
features at these wavelengths, however, their relative strengths as 
well as widths are markedly different in NGC\,4418. The 6.85\,$\mu$m
and 7.3\,$\mu$m band ratios are consistent with the features observed
towards Sgr A$^{*}$, which have been attributed to CH deformation
modes in HAC-like dust grains (Chiar et al. \cite{Chiar}). Thus, as
for the Galactic centre, both ice characteristics for shielded dense
molecular cloud environments and HAC-like grain characteristics for
diffuse ISM dust seem to be present along the line of sight. This
conclusion could be tested through observations in the 3\,$\mu$m
window, which contains the strong 3\,$\mu$m H$_{2}$O ice band and the
3.4\,$\mu$m CH stretching modes of HAC materials. Guided by variations
seen for the Galactic centre region (Chiar et al. \cite{Chiar}), and
between M\,82 and NGC\,1068 (Sturm et al. \cite{Sturm}), we speculate
that the relative weight of ice and HAC components may vary considerably
among galaxies.

\begin{table}[t]
\caption{Observed parameters for the NGC 4418 features.}
\begin{tabular}{lllll}
\hline
$\lambda$& $\tau$&  $\tau_{\rm int}$& carrier& N\\
$[\mu m]$ & &[cm$^{-1}$] & &[10$^{17}$~cm$^{-2}$]\\
\hline
6.0\hspace*{6mm} & 0.8\hspace*{6mm} & 87 & H$_{2}$O\hspace*{10mm} & 73 \\
6.85& 1.1 & 27 & HAC?     & $\sim$200\\
7.3 & 0.6 & 12 & HAC?     & $\sim$200\\
7.67& 0.12& 1.2& CH$_{4}$ & 1.6\\
9.7 & 7   & ---& silicates& ---\\
18  & $\sim$1.5& --- & silicates & ---\\
    &     &    & atomic H & $\sim$2$\times$10$^{6\,*}$ \\
\hline
\multicolumn{5}{l}{{$^*$} Column to the mid-IR dust source calculated assuming}\\ 
\multicolumn{5}{l}{Galactic gas to dust ratios. Note that the X-ray absorbing}\\
\multicolumn{5}{l}{column to the central engine may be even higher.}
\label{optdepths}
\end{tabular}
\end{table}

\subsection{Dust and ice column densities}

The column densities of the ice species can be derived by dividing
the integrated optical depth ($\tau_{\rm int}$) by the molecular 
band strength (Gerakines et al. \cite{Gerakines}; Boogert et al. \cite
{Boogert97}). Table\,\ref{optdepths} summarizes the computed column 
densities of CH$_{4}$ and H$_{2}$O ice. The column densities are 
consistent with those derived for embedded massive protostellar objects in 
molecular clouds (Boogert et al. \cite{Boogert96}; Keane et al. 
\cite{Keane}).
Also shown in Table\,\ref{optdepths} are the calculated HAC column 
densities for the 
6.85 and 7.3\,$\mu$m bands, assuming the intrinsic integrated band 
intensities for saturated aliphatic hydrocarbons from values by Wexler
(\cite{Wexler}).
These integrated intensities are stronger than other current literature 
values (Sandford et al. \cite{Sandford}; Furton et
al. \cite{Furton}). For NGC\,4418 a 
substantial fraction of the carbon is locked up in HAC ($\sim$20\%) as 
compared to the Galactic centre (a few \% in Sgr\,A$^{*}$; Pendleton
et al. \cite{Pendleton}).\\
\indent As mentioned previously (Sect.\,2), two continua, differing only
longward of 8\,$\mu$m were considered for NGC\,4418 (Fig.\,\ref{mirfir_sed}). 
The effect of the different continua on the silicate feature is noticable
when fitting the silicate profile by the Galactic Centre Source
GCS\,3, a pure absorption feature, i.e. no intrinsic emission
(Figer et al. \cite{Figer}). However, the exact optical depth is
still difficult to determine due to saturation of the silicate feature
in NGC\,4418. Adopting the dotted line in Figure\,\ref{mirfir_sed} 
results in a NGC\,4418 silicate profile in which the blue wing is well
matched but the red wavelength wing is poorly fitted by the GCS\,3
profile. If however the dashed line is adopted, then the fit to the
red wing improves while still maintaining a reasonable match to the
blue wing of the 10\,$\mu$m silicate band. The 9.7\,$\mu$m and
18\,$\mu$m optical depths for the second case are $\tau_{9.7}\sim$7 
(corresponding to A$_{\rm V}\sim$130) and $\tau_{18}\sim$1.5.
These numbers, along with the derived hydrogen column density, are 
shown in Table\,1. The apparent optical
depth ratio of the 9.7/18\,$\mu$m silicate bands, regardless of
adopted continuum, is significantly larger than the ratio determined
by Demyk et al. (\cite{Demyk}) for two Galactic protostars. This might
suggest that complex radiative transfer effects are important, which
are however beyond the scope of this Letter.

\section{Discussion and conclusions}

We have compared the ISO-PHT-S spectrum of NGC\,4418 with spectra of
our template sources and found no sign of PAH 
emission, neither from the nucleus, nor from that part of the disk 
contained within the 24$\arcsec\times$24$\arcsec$ ISO-PHT-S aperture.
Instead we found deep absorption features imposed upon a featureless 
mid-IR continuum. We identify the 6--8$\mu$m absorption features with 
foreground ices and HAC-like grains.\\
\indent The nature of the central source in NGC\,4418 cannot be
infered from the observed mid- to far-IR spectrum alone, given the
absence of any ``signposts'', like the 6.2\,$\mu$m PAH emission feature 
(F$_{\rm PAH}<$6$\times$10$^{-20}$\,W/cm$^2$), and of fine structure 
lines ($[$\ion{O}{i}$]$,
$[$\ion{C}{ii}$]$ and $[$\ion{O}{iii}$]$, Malhotra et al. (\cite{Malhotra});
F$_{\rm [\ion{Ne}{ii}]}<$2.3$\times$10$^{-20}$\,W/cm$^2$, from archival 
ISO-SWS data). Both a heavily enshrouded AGN or a similarly extincted 
nuclear starburst could be responsible for the observed continuum.
Even if a starburst could be accomodated within the compact central
source ($<$70\,pc in the mid-IR (Scoville et al. \cite{Scoville}) 
and 25--70\,pc at 6 and 20\,cm (Eales et al. \cite{Eales}; Kawara et al. 
\cite{Kawara90})), it would be highly unlikely to block the escape
of any mid-IR starburst indicator from a region of that size. The most
likely origin is therefore a heavily enshrouded AGN as suggested 
previously by Roche et al. (\cite{Roche86}), Kawara et al. (\cite{Kawara90}) 
and Dudley \& Wynn-Williams (\cite{DudleyWynn}). The 0.1$\arcsec$ point 
source (5mJy) detected with the Parkes Tidbinbilla Interferometer 
(PTI) at 13cm (Kewley et al. \cite{Kewley}; L.Kewley, priv.comm.) 
may actually pinpoint the AGN itself. Far-IR 
to millimetre sizes are less well constrained but the warm IRAS colours 
suggest that the emission in this range also arises in the nuclear region.\\
\indent Our finding of strong absorptions due to cold silicates and ices in 
NGC\,4418 leads us to believe that the same absorptions may be present
in the mid-IR spectra of other galaxies. Indeed, we have found similar
absorptions in about a dozen of 225 galaxies observed
spectroscopically by ISO. Spectral identifications
have to be done with great care since simultaneous presence of PAH
emission makes other spectra more complex than the one of NGC\,4418.\\
\indent Since the overall shape of the NGC\,4418 6--11\,$\mu$m spectrum 
with its maximum near 8\,$\mu$m mimicks at first glance a PAH spectrum,
we point out the need for high S/N spectra to clearly identify the 
indicators for bona fide PAH spectra or absorption dominated spectra. 
The most obvious discriminator is the 6.2$\mu$m PAH peak to be contrasted 
with the 6.5--6.7\,$\mu$m pseudo-maximum in absorption spectra, which is 
due to a window between two absorption features. In a forthcoming paper 
we will address this issue for our large ISO galaxy sample.

\begin{acknowledgements}
The authors wish to thank Olivier Laurent and Leticia Mart\'{\i}n-Hern\'{a}ndez 
for performing the ISO-CAM and ISO-LWS data reduction, respectively, as 
well as Ilse van Bemmel, Peter Barthel, Matthew Lehnert, Dave Sanders, 
Nick Scoville, Ralf Siebenmorgen, Eckhard Sturm and Dan Tran for discussions.
\end{acknowledgements}


\begin{thebibliography}{}

\bibitem[1996]{Boogert96}Boogert A.C.A., Schutte W.A., Tielens A.G.G.M., 
        et al., 1996, A\&A 315, L377
\bibitem[1997]{Boogert97}Boogert A.C.A., Schutte W.A., Helmich F.P., 
        et al., 1997, A\&A 328, 649
\bibitem[2000]{Chiar}Chiar J.E., Tielens A.G.G.M., Whittet D.C.B.,
        et al., 2000, ApJ 537, 749
\bibitem[2000]{Clavel}Clavel J., Schulz B., Altieri B., et al., 2000,
        A\&A 357, 839
\bibitem[1990]{Condon}Condon J.J., Helou G., Sanders D.B, Soifer B.T.,
        1990, ApJS 73, 359
\bibitem[1999]{Demyk}Demyk K., Jones A.P., Dartois E., et al.,
        1999, A\&A 349, 267
\bibitem[1997]{Dudley97}Dudley C.C., 1997, Ph.D. thesis, Univ. Hawaii
\bibitem[1997]{DudleyWynn}Dudley C.C., Wynn-Williams C.G., 1997, 
        ApJ 488, 720
\bibitem[1990]{Eales}Eales S.A., Becklin E.E., Hodapp K.-W., et al.,
        1990, ApJ 365, 478
\bibitem[1999]{Figer}Figer D.F., McLean I.S., Morris M., 1999, ApJ
        514, 202
\bibitem[1999]{Furton}Furton D.G., Laiho J.W., Witt A.N., 1999, ApJ
        526, 752
\bibitem[1998]{Genzel}Genzel R., Lutz D., Sturm E., et al., 1998, 
        ApJ 498, 579
\bibitem[1995]{Gerakines}Gerakines P.A., Schutte W.A., Greenberg J.M.,
        van Dishoeck E.F., 1995, A\&A 296, 810
\bibitem[2000]{Gibb}Gibb E.L., Whittet D.C.B., Schutte W.A., et al., 
        2000, ApJ 536, 347
\bibitem[2000]{Helou}Helou G., Lu N.Y., Werner M.W., Malhotra S.,
        et al., 2000, ApJ 532, L21
\bibitem[1989]{Kawara89}Kawara K., Nishida M., Phillips M.M., 1989,
        ApJ 337, 230
\bibitem[1990]{Kawara90}Kawara K., Taniguchi Y., Nakai N., et al., 
        1990, ApJ 365, L1
\bibitem[2000]{Keane}Keane J.V., Tielens A.G.G.M., Boogert A.C.A., 
        et al., 2000, A\&A, Submitted
\bibitem[2000]{Kewley}Kewley L.J., Heisler C.A., Dopita M.A, et al.,
        2000, ApJ 530, 704
\bibitem[1995]{Lehnert}Lehnert M.D., Heckman T.M., 1995, ApJS 97, 89 
\bibitem[2000]{Lisenfeld}Lisenfeld U., Isaak K.G., Hills R., 2000,
        MNRAS 312, 433
\bibitem[1998]{Lutz}Lutz D., Spoon H.W.W., Rigopoulou D., et al., 
        1998, ApJ 505, L103
\bibitem[1999]{Malhotra}Malhotra S., et al.,
        1999, In: The Universe as seen by ISO, P. Cox, M.F. Kessler 
        (eds.), ESA-SP 427, 813
\bibitem[1994]{Pendleton}Pendleton Y.J., Sandford S.A., Allamandola
        L.J., et al., 1994, ApJ 437, 683
\bibitem[1994]{Ridgway}Ridgway S.E., Wynn-Williams C.G., 1994, 428, 609
\bibitem[1999]{Rigopoulou}Rigopoulou D., Spoon H.W.W., Genzel R., et al., 
        1999, AJ 118, 2625
\bibitem[1986]{Roche86}Roche P.F., Aitken D.K., Smith C.H., James S.D., 
        1986, MNRAS 218, 19P
\bibitem[1993]{Roche93}Roche P.F., Chandler C.J., 1993, MNRAS 265, 486
\bibitem[1991]{Sandford}Sandford S.A., Allamandola L.J., Tielens
        A.G.G.M., et al., 1991, ApJ 371, 607
\bibitem[2000]{Scoville}Scoville N.Z, Evans A.S., Thompson R., et al.,
        2000, AJ 119, 991
\bibitem[1989]{Soifer}Soifer T., B\"ohmer L., Neugebauer G., et al.,
        1989, AJ 98, 766
\bibitem[2000]{Spoon}Spoon H.W.W., Koornneef J., Moorwood A.F.M., et al.,
        2000, A\&A 357, 898
\bibitem[2000]{Sturm}Sturm E., Lutz D., Tran D., et al., 2000, A\&A 358, 481
\bibitem[2000]{Tran}Tran Q.D., Lutz D., Genzel R., et al., ApJ, Submitted
\bibitem[1967]{Wexler}Wexler A.S., 1967, Applied Spec. Rev. 1, 29
\bibitem[1996]{Whittet}Whittet D.C.B., Schutte W.A., Tielens
        A.G.G.M., et al., 1996, A\&A 315, L357
\bibitem[1993]{Wynn}Wynn-Williams C.G., Becklin E.E., 1993, ApJ 412, 535
\end{thebibliography}
\end{document}